\documentclass[fleqn,10pt]{wlscirep}

\usepackage{breakurl}
\usepackage{adjustbox} 
\usepackage{graphicx}
\definecolor{col1}{rgb}{0.87, 0.36, 0.51} 
\definecolor{aqua}{rgb}{0.0, 1.0, 1.0}
\definecolor{darkorchid}{rgb}{0.6, 0.2, 0.8}
\definecolor{fuchsia}{rgb}{1.0, 0.0, 1.0}
\definecolor{lightblue}{rgb}{0.12, 0.56, 1.0}
\definecolor{darkgreen}{rgb}{0.01, 0.75, 0.24}
\definecolor{cerise}{rgb}{0.85, 0.2, 0.53}
\definecolor{lavanda}{rgb}{0.96, 0.73, 1.0}
\definecolor{debianred}{rgb}{0.84, 0.04, 0.33}

\title{Networks of plants: how to measure similarity in vegetable species}

\author[1]{Gianna Vivaldo}
\author[2]{Elisa Masi}
\author[2]{Camilla Pandolfi}
\author[2]{Stefano Mancuso}
\author[1,3,4,*]{Guido Caldarelli}

\affil[1]{IMT School for Advanced Studies, Piazza San Francesco 19, 55100 Lucca, Italy.}
\affil[2]{Universit\`a di Firenze, Dipartimento di Scienze Produzioni Agroalimentari e dell'Ambiente (DISPAA)
Viale delle Idee, 30 50019 Sesto Fiorentino Firenze.}
\affil[3]{London Institute for Mathematical Sciences, 35a South St. Mayfair W1K 2XF London UK.}
\affil[4]{Istituto dei Sistemi Complessi (ISC), Roma, Italy.}
\affil[*]{Guido.Caldarelli@imtlucca.it}


\begin{abstract}
Despite the common misconception of nearly static organisms, plants do interact continuously with the environment and with each other. It is fair to assume that during their evolution they developed particular features to overcome problems and to exploit possibilities from environment.  In this paper we introduce various quantitative measures based on recent advancements in complex network theory that allow to measure the effective similarities of various species. By using this approach on the similarity in fruit-typology ecological traits we obtain a clear plant classification in a way similar to traditional taxonomic classification. This result is not trivial, since a similar analysis done on the basis of  diaspore morphological properties  do not provide any clear parameter to classify plants species. Complex network theory can then be used in order to determine which feature amongst many can be used to distinguish scope and possibly evolution of plants. Future uses of this approach range from functional classification to quantitative determination of plant communities in nature.
\end{abstract}

\begin{document}
\flushbottom
\maketitle
\thispagestyle{empty}

\section{Introduction} \label{sec:intro}
Plants are the building blocks of food production on Earth. Their role is crucial in the transformation and use of chemical energy to sustain the energy transfer in food webs and ultimately to feed any animal species. Despite their importance, they are seldom considered in ecological analysis of food webs and more generally they have attracted a relatively small interest for people studying complex networks. Actually, the study of vegetable world is revealing more and more evidence of the fact that different plants have many unexpected ties connecting them with each other. For instance,  they are able to interact with the environment and to actively defend themselves from predators. On this respect, we note that the recipe that animal developed for the same purpose was to create an energetically expensive neural and locomotion system. This is mainly due to the fact that single individuals are 
indeed ``in-dividual", that is they cannot be divided without killing them. Plants ``individuals" instead can even propagate by their division and generally tolerate a loss of some of their parts. For this reason, in order to perform defensive tasks, they developed a series of features remarkably different than those of animal species. As a result,  plants communicate, or ``signal," with each other, using a complex internal analysis system to find nutrients, spread their species and even defend themselves against predators \cite{Brenner2006413,baluska2007}. Plants have solved all these problems in different ways, shaping the plant growth, adapting to the different environmental constrains, using different kind of vectors in many phase of their life to overcome their immobility. One of the most critical stages in the life of any plant is the dispersal of seeds into a suitable habitat. To do this plants make effective use of many external agents such as wind, water, insects or higher animals. 
In order to track the many different series of strategies we need a measure to determine how much the same feature (i.e. fruit shape or diaspora mechanism) are different in two distinct species. A similar process  is at the basis of taxonomic classification where plants are clustered according for example to properties (number of stamen) of plants, while cladistic classification is instead based on common ancestory. 

In this paper we use network analysis\cite{caldarelli2007scale,Boccaletti2006175,barrat2004architecture} of some relational data about different plants with the aim of finding classes of ``similar'' plants. This analysis allows a clustering of species able to reveal and quantify similarity with respect to different species. In network theory, the various elements of an ensemble (i.e. plants in vegetal kingdom) are represented by vertices and they can be joined by considering common features they have. The number of common edges becomes then a quantitative proxy of relationships that are otherwise impossible to measure. In this respect this is similar to what happen in technological systems where the number of e-mails\cite{caldarelli2004preferential}, likes on Facebook\cite{zollo2015emotional}, or retweet between two persons\cite{eom2015twitter}, becomes a number assessing the strength of an acquaintance or even friendship.  
When passing to biology, network theory has been fruitfully used to determine structure and robustness of Food Webs\cite{Dunne01102002}, as well as the structure of protein interactions in the cell\cite{Stelzl2005957} with important applications to human diseases\cite{leecomorbidity}.  As previously mentioned, compared to other topics in biology, plants received a minor attention from networks scientists, despite some tentatives of comparing different ecosystems looking for steady (i.e. ``universal'') behaviours\cite{caretta2008}.  In order to adapt to the environment in which they live, plants have evolved an astonishing number of different mechanisms and structures to disperse their seeds. Typically,  plants evolved in time to adapt to the environment in which they lived, so that only the mutations giving a comparative advantage with others were selected. Today, after $500$ million years of plant evolution we are witnessing a huge differentiation in the features of plants as seed form and dispersal structure. Of the $250,000$ today known flowering plants just a small fraction ($5000$) has been classified in available databases on the basis of the variety of seed features. 

\begin{figure}[h!]
\centering
\includegraphics[width=0.6\textwidth]{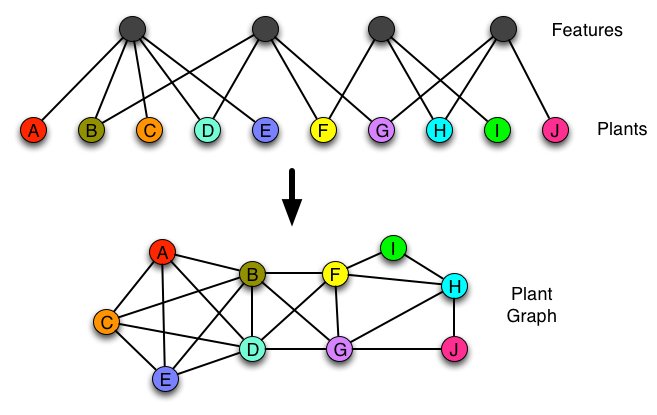} 
\caption{\textbf{Bipartite network structure.} From the original graph one can create a graph made by only one of the two sets.} 
\label{fig1}
\end{figure}

These features can be represented by a graph of correlation, providing an effective taxonomy of vegetable species. The basic idea is to represent the information on plants, by means of a bipartite graph. A graph $G(N,E)$ is a mathematical object composed by $N$ vertices and $E$ edges. In a bipartite graph, vertices are divided in two sets, and the connections are made only from vertices of one set towards vertices of the other set. From one side we have the different plants, on the other side the various features. 
This information is transformed into two other graphs made by vertices of the same kind (see Fig.\ref{fig1}). In the first case we connect plants with plants on the basis of their common features. In the second case we connect features with features on the basis of how many plants have similar behaviour. Community detection\cite{raghavan2007near} in such a graph are a powerful method to classify in a quantitative way the different vertices creating a taxonomic tree\cite{newman2004finding}.

We present here the main results on the analysis conducted on the datasets considered; further detailed analysis is present in the Supplementary Information provided with this paper.

\section{Results} \label{sec:results}
The results presented here are computed on the dataset D3 Dispersal and Diaspore Database\cite{hintze2013d3} suitably represented as a network as shown in Fig.\ref{fig1} and with  the details presented in the section ``Data''.
\subsection{Basic network analysis.}
Plants species networks $G^{P}$ are defined by considering as vertices the plant species $i$ and $j$ in the database; two vertices are linked if they share at least one common property. The $2,662$ plants species analyzed are representative of 111 families, but the dataset is not homogeneous in terms of families percentages, being dominated by \textit{Asteraceae} ($12.81 \%$), \textit{Poaceae} ($8.72 \%$), \textit{Cyperaceae} ($5.63 \%$), \textit{Brassicaceae} ($5.41 \%$), \textit{Rosaceae} ($5.33 \%$), and \textit{Fabaceae} ($4.58 \%$).  
In the following we consider both properties related to diaspora morphology ($G^{P}_1$) as well as fruit typology ($G^{P}_2$).
For the various networks, we considered {\bf size} (number of edges), {\bf measure} (number of vertices), {\bf degree} (average and its distribution), {\bf density} (the ratio of actual vertices against the possible ones), {\bf clustering} and finally (in the next section) the {\bf community structure}.

\subsubsection{Diaspora-based graph}
 A weight $w_{ij}$ of each link $e_{ij}$ can be defined by the total number of shared properties between plant $i$ and plant $j$. The measure of $G^{P}_1(N,E)$ is given by  $N=2,662$ vertices (plants species) and the size by $E = 1,176,968$ edges. The maximum and minimum number of properties shared by two plants are equal to $1$ and $4$, respectively. The $69.84 \%$ of plants share one property, only, and the proportion of edges with weight $w_{ij}=1$ represents the $89.47\%$ of $E$. On the contrary, just the $3.2 \%$ of the species share four properties, and $w_{ij}=4$ links accounts for the $0.1\%$ of the graph total number of edges $E$.

\begin{figure}[!h]
\centering
\includegraphics[width=0.8\textwidth]{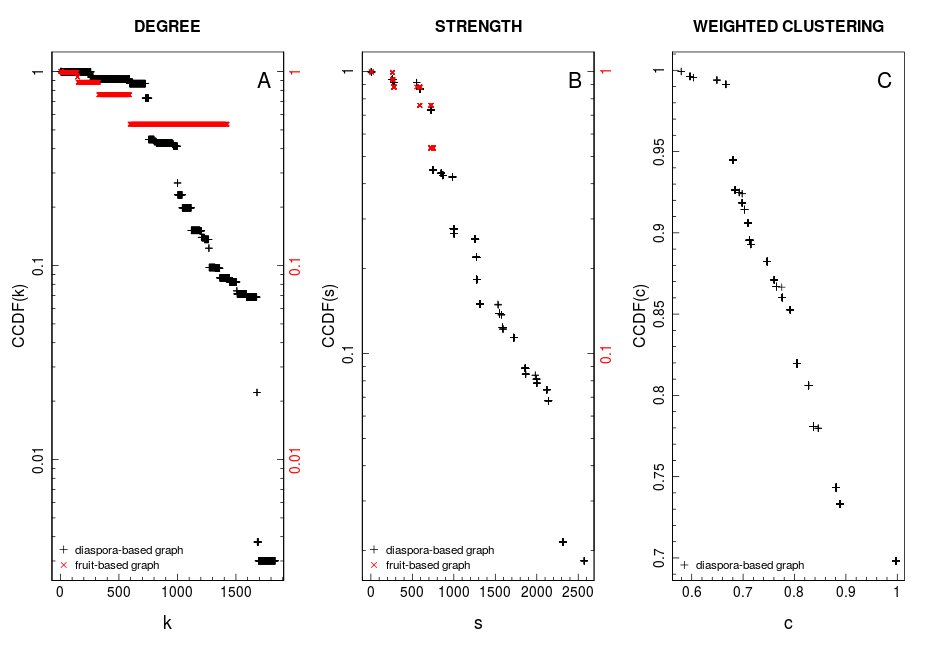} 
\caption{\textbf{Basic network analysis.} Complementary cumulative distribution functions (CCDF) of degree and strength are reported in log-line scale in panel A and B, respectively, for both diaspora-based network (black crosses) and fruit-based network (red crosses). Panel C, moreover, shows $G^P_1(N,E)$ weighted clustering coefficient distribution. More precisely, CCDF (on y-axis) is plotted versus the weighted clustering parameter (x-axis) on linear scale.} 
\label{fig2}
\end{figure}

As regards the basic metrics,  we can describe $G^{P}_1$ as a weakly connected graph, whose density $\frac {2E} {N(N-1)}$ is equal to $0.332$, the global weighted clustering coefficient is $0.84$, and the nodes mean degree is $\overline{k}=\frac 1 N \sum_{i=1,N} k_i=\frac {2E}{N} = 884.27$. 
The network degree distribution $P(k)$, representing the fraction of vertices with degree $K > k$, is shown in Fig.~\ref{fig2} (panel A, black crosses). More in details, the log-line plot displays  $G^{P}_1$ degree complementary cumulative distribution function (CCDF). Analogously, panel B (black crosses) displays the graph strength distribution, where the vertices strength $s$ takes into account their connections total weight. Besides, panel C shows $G^{P}_1$ local clustering coefficient, defined as the tendency among two vertices to be connected if they share a mutual neighbour. 
Taken as a whole, Fig.~\ref{fig2} suggests that plants network is not dominated by some central nodes with a huge amount of connections linking them to all the other minor vertices. 

\subsubsection{Fruit-based graph}
We extended our analysis to the ecological properties of the fruit related to seed dispersal. Following the same approach, we created  $G^P_2(N,E)$ as a projection of the bipartite graph where the plants are associated to fruit features. This creates a graph made up of $N=2,662$ vertices (plants species) connected by $E=1,265,831$ edges. 

Also this graph is sparse with a density $0.357$ and an average degree equal to $\overline{k}=951.04$. The weight $w_{ij}$ of each link $e_{ij}$ is given by the total number of shared properties between plant $i$ and plant $j$. The maximum number of properties shared by two plants is one, thus suggesting how fruit typology is a more strict parameter to classify plants behaviour related to diaspores, since plants cannot share more than a single trait. Moreover the properties are \textit{mutually exclusive}, i.e. each species possesses just one of the eight properties analyzed. That can be easily verified by building the bipartite projection of the fruit typology graph (not shown) made up by eight vertices, each one equal to a fruit typological property. The number of links of such a network is zero, meaning that two different properties do not share any species between them.
Figure~\ref{fig2} (panel A, red crosses) shows the fruit-based graph degree CCDF by log-line scale, while $G^P_2$ strength CCDF is displayed in panel B (red crosses). The weighted clustering coefficient distribution is not shown for that second graph since  $G^P_2$ is made up by fully connected isolated subgraphs, apart for a couple of nodes. Thus the local  clustering coefficient is equal to $1$ for all the vertices, while it is undefined for the two interconnected nodes (for a more deep description of the analyzed network metrics, refer to Methods section). 

\subsection{Community detection analysis.} 
\subsubsection{Diaspora-based graph}
\begin{figure}
\centering
\includegraphics[width=0.8\textwidth]{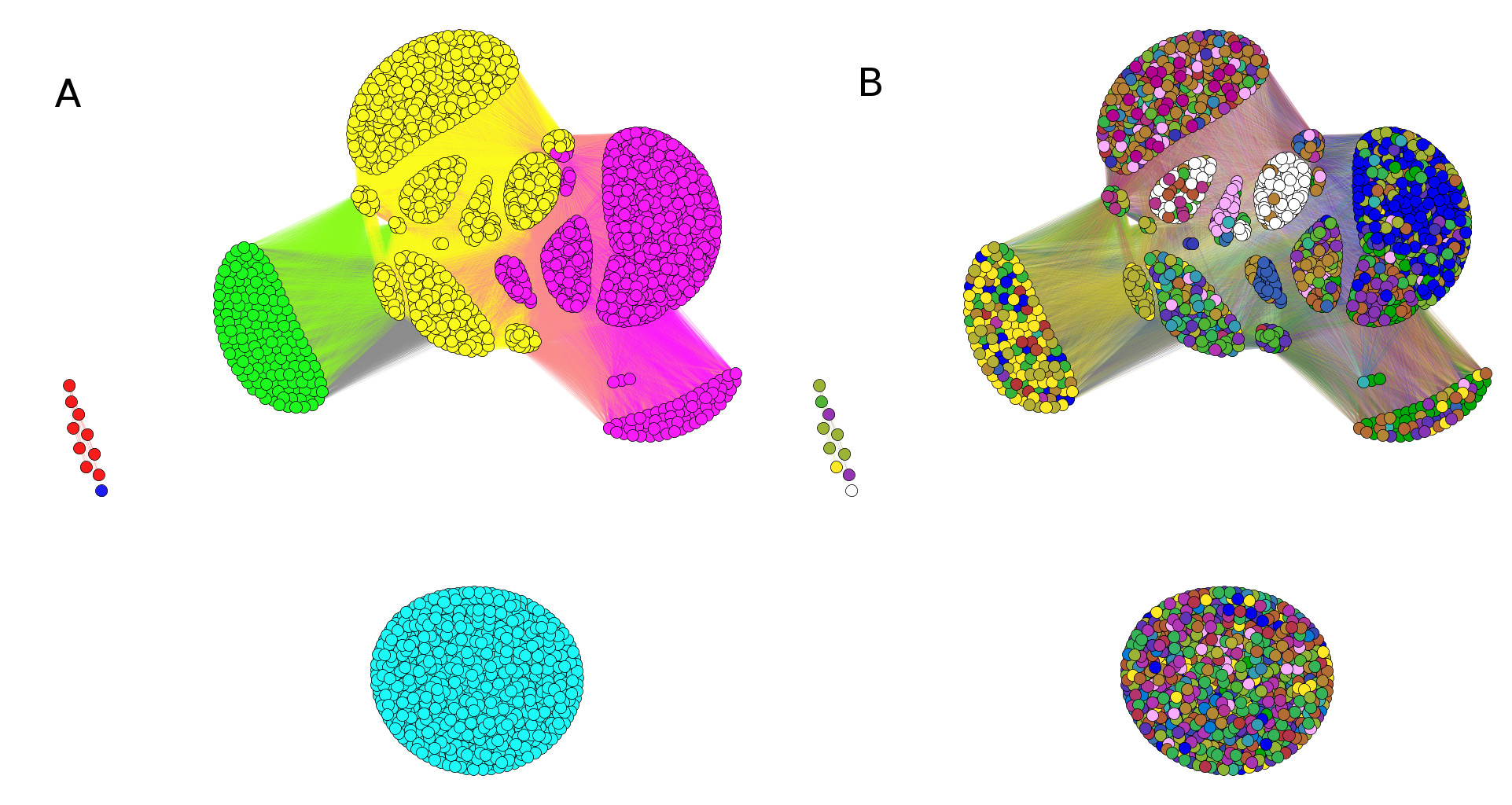} 
\caption{\textbf{Communities detection based on diaspore morphology.} The graphs refers to $G^P_1(N,E)$ communities detection by modularity method. Panel A shows the six communities which are detected: green, yellow, and fuchsia communities are highly connected components. On the contrary, red, blue and cyan clusters are isolated components. While cluster blue just embeds a single species (\textit{X Calammophila baltica Brand}), cluster cyan is quite big, being composed by the $28.29 \%$ of total species present in the database $D^{3}$, for a total of $12$ different families. Panel B shows the families belonging to each cluster. \textit{Asteraceae} (blue, $12.81\%$), \textit{Poaceae} (white, $8.72\%$), \textit{Cyperaceae} (dark green, $5.63\%$), \textit{Brassicaceae} (yellow, $5.41\%$), \textit{Rosaceae} (cerise, $5.33\%$) are some of the most numerous. The heterogeneous distribution of families inside each clusters is evident.} 
\label{fig3}
\end{figure}

We show the result of the community detection on the first graph $G^{P}_1$ in Table~\ref{tab:diaapp_clusters}. The communities detection results are obtained by using different algorithms: (i) fastgreedy (FG), (ii) walktrap (WT), (iii) Blondel's modularity optimisation algorithm (BL) and (iv) label propagation (LP) (see Methods). 
Each line corresponds to a different subgraph, i.e. a filtered-by-edges-weight versions of $G^{P}_1$, with $w_{ij} \in [1,2,3,4]$.
\begin{table}
\centering
\begin{adjustbox}{max width=0.6\textwidth}
\begin{tabular}{|c|c|c|c|c|c|c|c|c|c} 
\hline
\hline
FG & WT  & BL & LP & weight & E & N & is.connected & density \\ \hline
5  &  6  &  6 &  6 &  1 &  1176968 & 2662 & FALSE  & 0.3323087 \\
4  &  7  &  6 &  2 &  2 & 123939 &  803 & TRUE &  0.3849001 \\
6  &  9  &  7 &  6 &  3 & 27009  &  343 & FALSE & 0.460488 \\
4  &  4  &  4 &  4 &  4 & 1395   &   85 & FALSE & 0.3907563 \\     		
\hline
\hline			
\end{tabular}
\end{adjustbox}
\caption{Plant species in diaspora-based plant graph are grouped on the basis of the common diaspore morphological properties. Four distinct communities detection algorithms were employed: FG = fastgreegy, WT =  walktrap algorithms, BL = Blondel modularity optimization, LP = label propagation. Four filtered-by-edges-weight versions of the graph were analyzed (one for each row). Graph edges weight integer values range from $1$ to $4$.}
\label{tab:diaapp_clusters}
\end{table}
Figure~\ref{fig3} shows the six communities detected by modularity algorithm (BL) in graph $G^P_1$. Colours refer both to cluster (panel A) and to families (panel B) membership. Looking to panel A, clusters 3 (cyan), 5 (red), and 6 (blue) are isolated components. The three bigger clusters and the corresponding families they embed are reported in Supplementary Information.
Such communities are not homogeneous in terms of family composition (see panel B). Hereafter, the composition of every cluster is summarised, together with the morphological properties that the element families share each other. Notice that one property can be shared by more than a single species in the same cluster, since diaspore morphological features are not mutually exclusive. 

\begin{itemize}
\item {\bf{cluster 1}}: $884$ species ($33.21 \%$ of database $D^{3}$ total species); prevailing families: \textit{Poaceae}, \textit{Fabaceae}, \textit{Rosaceae}, \textit{Plantaginaceae}, \textit{Polygonaceae}. $709$ species have \textbf{nutrient} diaspores, followed by $447$ showing \textbf{flat/wings} diaspore morphology; $204$ times is encountered the \textbf{elongated} feature.
\item {\bf{cluster 2}}: $858$ species ($32.23 \%$) dominant families: \textit{Asteraceae}, \textit{Cyperaceae}, \textit{Ranunculaceae}, \textit{Rosaceae}, \textit{Apiaceae}, \textit{Amaranthaceae}, \textit{Salicaceae}, \textit{Caprifoliaceae}, \textit{Potamogetonaceae}. The vast majority of the species ($782$) show \textbf{elongated} diaspore trait; other common observed properties are: \textbf{hooked} ($220$), \textbf{ballo/aerenchym} ($224$), and \textbf{flat/wings} ($140$).
\item {\bf{cluster 3}}: $753$ species ($28.29 \%$), sharing property \textit{no specialization}. Notwithstanding its big dimensions, that cluster is a completely isolated component robust to changes in clustering algorithms. The leading families belonging to this cluster  are summarized in Supplementary Information. They all share the same \textbf{no specialization} property concerning diaspore morphology. That category refers to species whose diaspores can have either a structured surface and no further appendages or specializations (e.g. many \textit{Caryophyllaceae}), or a smooth surface and no further appendages or specializations (e.g. many \textit{Brassicaceae}). \textit{Caryophyllaceae} and  \textit{Brassicaceae} are two of the most numerous families with $86$ and $43$ species each respectively, besides \textit{Orchidaceae} ($61$) and \textit{Orobanchaceae} ($48$). Many species found in this cluster are characterized by very small, dust-like seeds, whose dispersal is easily achieved through the wind movements, even without specialized structures.
\item {\bf{cluster 4}}: $157$ species ($5.9 \%$); prevailing families: \textit{Brassicaceae}, \textit{Juncaceae}, \textit{Plantaginaceae}, \textit{Asteraceae}, \textit{Lamiaceae}. All these species share \textbf{mucilaginous} diaspore property. 
\item {\bf{cluster 5}}: 9 plants species belonging to \textit{Hydrocharitaceae}, \textit{Brassicaceae}, \textit{Polygonaceae}, and \textit{Araceae} families. They all show \textbf{other specialization} concerning diaspore morphology. More in detail, 7 out of 9 are aquatic plants (5 species of \textit{Hydrocharitaceae} and 2 of \textit{Araceae} family); 1 species belongs to \textit{Brassicaceae} and 1 to \textit{Polygonaceae}. The 5 species of \textit{Hydrocharitaceae} are strictly related: like other \textit{Hydrocharitaceae}, they are aquatic plants that release their diaspore in water and that, conversely to other plants of the same family, have seeds with very low nutrients content; more, they do not set seeds regularly, preferring asexual reproduction; in both cases (sexual or asexual reproduction) water movements allow the dispersal; the 2 other aquatic (\textit{Araceae}) also prefer asexual reproduction; having no or little roots, the whole plants can float and disperse; the species belonging to the family of \textit{Brassicaceae} has dehishent fruits; finally, the species of \textit{Polygonaceae} rarely produces viable seeds and reproduction is normally asexual (by bulbils).
\item {\bf{cluster 6}}: 1 isolated plant, \textit{X Calammophila baltica Brand (Poaceae)} which doesn't show any of the used morphological properties with the other species. 
\end{itemize}

The total number of species which are part of each cluster, and the corresponding total number of families to which they belong are shown in Tab.~\ref{tab:test1_species_fam}. Notice the persistent heterogeneity of each cluster. The percentage reported in the third column of Tab.~\ref{tab:test1_species_fam} refer to the relative number of species inside each cluster with respect to the total number of species present in $D^{3}$ database ($2,662$). Analogously, the relative number of families inside each cluster (last column, Tab.~\ref{tab:test1_species_fam}), is referred to the total amount of families inside the dataset, i.e. $111$. Each plant belongs to a single cluster, while different families can characterize different clusters.
\begin{table}
\centering
\resizebox{0.4\textwidth}{!}{%
\begin{tabular}{|c|c|c|c|c|} 
\hline
\hline
\textbf{cluster} & \textbf{species} & \textbf{$\%$} & \textbf{families} & \textbf{$\%$}\\
\hline
\hline
1 & $884$ & $33.21 \%$ & $73$ & $65.76 \%$\\
2 & $858$ & $32.23 \%$ & $44$ & $39.64 \%$\\
3 & $753$ & $28.29 \%$ & $57$ & $51.35 \%$\\
4 & $157$ & $5.9 \%$ & $12$ & $10.81 \%$\\
5 & $9$ & $0.34 \%$ & 4 & $3.6 \%$\\
6 & $1$ & $0.04 \%$ & 1 & $0.9 \%$ \\ 
\hline
\hline
\end{tabular}}
\caption{$G_{1,P}(N,E)$ clusters composition on the basis of diaspore morphological properties. The total number of species corresponds to the measure $N=2,662$ of the graph. The total number of families is equal to $111$. Species and family percentage are referred to that values.}
\label{tab:test1_species_fam}
\end{table}
The results are generally robust to changes in the detection algorithm, and to sizes of the filters employed over edges weights. In general we note that the network $G^P_1$ is made up of a small number of clusters. Some of them behave like weakly-connected components that can be split into a different number of sub-clusters, depending on the applied methodology. For this reason we also made the same analysis on a filtered versions of $G^{P}_1$  to better focus on the largest components.

\begin{figure}
\centering
\includegraphics[width=0.8\textwidth]{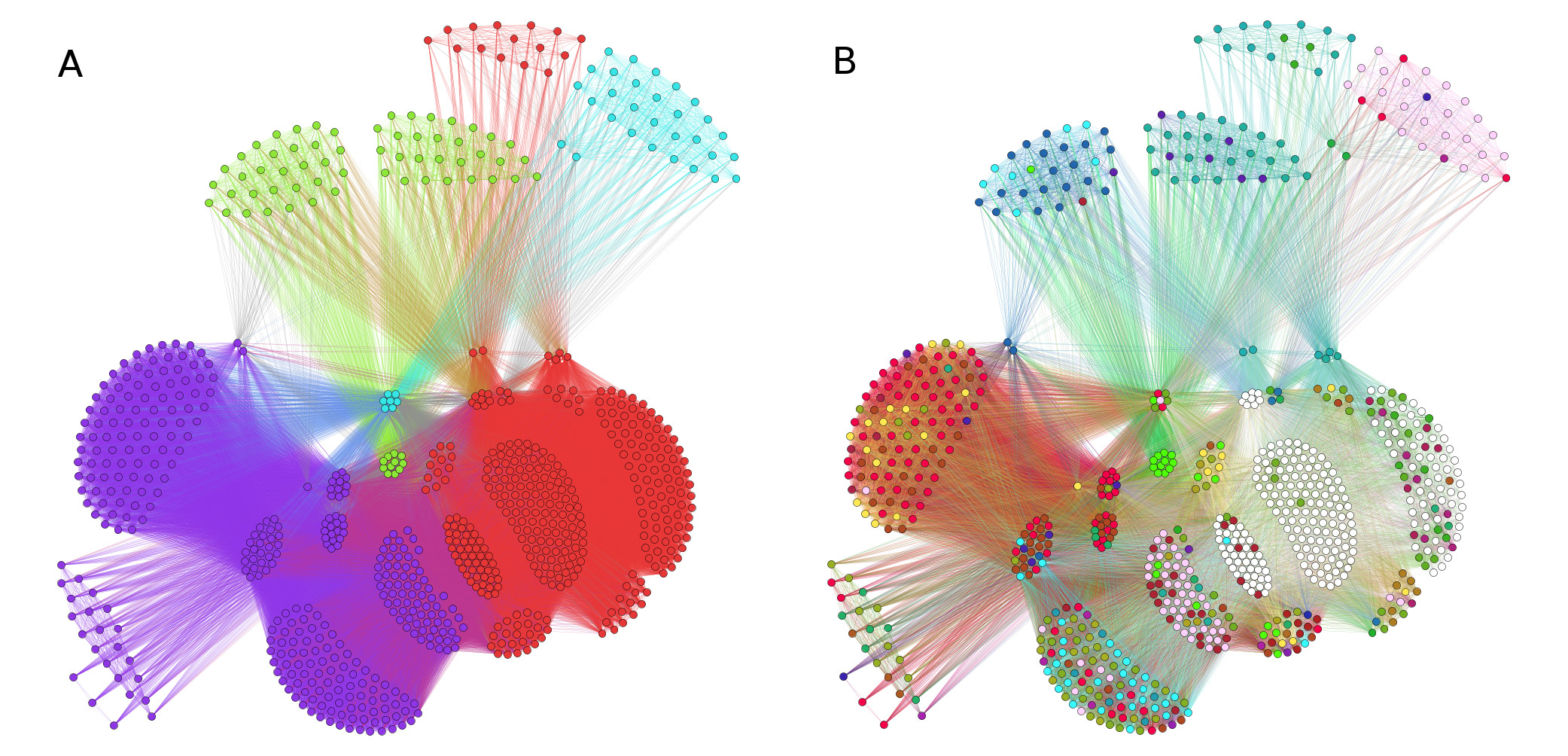} 
\caption{\textbf{Communities detection on a filtered version of $G^P_1(N,E)$ graph.} In that case, edges with weight $w_{ij} = 1$ are removed from the original graph. Four clusters are detected. Clearly each cluster is highly heterogeneous in terms of families composition, but more correspondences are found, and some families begin to dominate some cluster (especially red and cyan clusters of left panel). Prevailing families are visible in panel B: \textit{Poaceae} (white), \textit{Cyperaceae} (red), \textit{Rosaceae} (cerise).} 
\label{fig4}
\end{figure}

{\bf Communities after pruning of Diaspora-based graph}
The same modularity analysis was performed on three filtered-by-edges-weight versions of the seeds features graph. 
Figure~\ref{fig4} shows the four communities detected by BL algorithm, after filtering by edges weight $w_{ij} > 1$, thus retaining plants connected by more that a single property.
In that way only $N=803$ vertices/plants species organized into $46$ families and $123,939$ links survive the pruning. Colors here keep the same meaning of Fig.~\ref{fig3}, so that each color in the right panel corresponds to one of the $46$ families present in the filtered dataset.

Again, detected communities are not homogeneous in terms of family composition. Anyway, more correspondences can be observed between the two panels of Fig~\ref{fig4}. Cluster 1 (red) and clusters 3 (cyan), for example, are less heterogeneous, being composed by \textit{Poaceae}  and \textit{Rosaceae} families, respectively (white and cerise dots in the right panel). 
Table~\ref{tab:test1_species_fam_w1} reports species and families amount and the corresponding percentage present in each cluster.
\begin{table}
\centering
\resizebox{0.4\textwidth}{!}{%
\begin{tabular}{|c|c|c|c|c|} 
\hline
\hline
\textbf{cluster} & \textbf{species} & \textbf{$\%$} & \textbf{families} & \textbf{$\%$}\\
\hline
\hline
1 & $352$ & $43.84 \%$ & $31$ & $ 27.9\%$\\
2 & $345$ & $42.96 \%$ & $27$ & $24.32 \%$\\
3 & $37$ & $4.61 \%$ & $7$ & $6.3 \%$\\
4 & $69$ & $8.59 \%$ & $7$ & $6.3 \%$\\
\hline
\hline
\end{tabular}}
\caption{Families and species composition for each cluster detected by BL method on a filtered version of $G^P_1$ graph ($w_{ij} > 1$). After filtering just $N=803$ vertices survive, corresponding each one to a different plant species. The total number of families is equal to $41$. Families percentage is referred to the total amount of families into the dataset ($111$).}
\label{tab:test1_species_fam_w1}
\end{table}
 A brief description of the four clusters identified by BL method is the following.
\begin{itemize}
\item {\bf{cluster 1}}: $352$ species ($43.84 \%$ of database $D^{3}$ total species); \textit{Poaceae} with $228$ species are clearly the prevailing family: see white nodes in panel B of Fig.~\ref{fig4}. They are followed by \textit{Juncaceae} ($14$ plants), \textit{Fabaceae}, \textit{Santalaceae}, \textit{Caprifoliaceae}, \textit{Pinaceae}. 
\\ All these species share that common properties: \textbf{nutrients} ($315$), \textbf{flat/wings} ($312$), \textbf{elongated} ($240$). They do not show (almost most of them)  \textbf{ballo/aerenchyms} and \textbf{mucilaginous} surfaces;
\item {\bf{cluster 2}}: $345$ species ($42.96 \%$); dominant families: \textit{Cyperaceae} ($89$), \textit{Rosaceae} ($48$), \textit{Ranunculaceae}($42$), \textit{Asteraceae}($29$). \textit{Cyperaceae} are visible as red dots in Fig.~\ref{fig4} (panel B) in the position corresponding to violet cluster of panel A. That cluster embeds species joined by \textbf{elongated} ($317$) and \textbf{hooked} ($211$) diaspores shape. \textbf{Ballo/aerenchyms} and \textbf{flat/wings} are shared by $175$ and $112$ species, respectively. Just $4$ species shows \textbf{mucilaginous} surfaces;
\item {\bf{cluster 3}}: $37$ species ($8.95 \%$); \textit{Rosaceae} family dominates with $23$ species, visible as cerise vertices in Fig.~\ref{fig4} (panel B) in the position corresponding to cyan cluster in panel A. Almost all of them share clearly two properties: \textbf{nutrients} and \textbf{ballo/aerenchyms} surfaces;
\item {\bf{cluster 4}}: $69$ species ($4.61 \%$), dominated by those belonging to \textit{Potamogetonaceae} ($20$), \textit{Plantaginaceae} ($19$), and \textit{Amaranthaceae} ($12$) families. All the species have \textbf{mucilaginous} surfaces, some of them show \textbf{flat} diaspores ($39$), in particular species belonging to \textit{Plantaginaceae} and \textit{Juncaceae} families; other individuals show \textbf{elongated} diaspore ($41$), especially \textit{Amaranthaceae}, \textit{Asteraceae}, \textit{Potamogetonaceae}.
\end{itemize}

Notice that after pruning $G^P_1$, the species dataset reduces to $803$ species/vertices and it is made up especially of \textit{Poaceae} ($28.39\%$), \textit{Cyperaceae} ($11.96\%$), and  \textit{Rosaceae} ($9.09 \%$). Different clusters are dominated by different families: \textit{Poaceae} (cluster $1$), \textit{Cyperaceae} (cluster $2$), and \textit{Rosaceae} (dominant family in cluster $3$, and second dominant family in cluster $2$).

In any case, some general conclusions can be drawn after pruning $G^P_1$.
\textit{Poaceae} family dominates cluster $1$ with $228$ species. This is a robust result, since before filtering out plants sharing a single property, \textit{Poaceae} were rather well grouped into a single cluster. 
\textit{Cyperaceae} family is present in cluster $4$ with $89$ species. Before pruning, that family was already one of the most copious in cluster $2$ with $134$ species, after \textit{Asteraceae}. On the contrary, \textit{Asteraceae}, which previously were copious (dominant family with $279$ species in cluster $2$, i.e. the magenta cluster in Fig.~\ref{fig3} (panel A )), now are quite disappearing, and just a thirty of them survive. The same happens for \textit{Caryophyllaceae}, which go from a hundred of species to no one taxa surviving the pruning. \textit{Rosaceae} family is present in cluster $2$ with $48$ species, and in cluster $3$ with $23$ species. Two single species belongs to cluster $1$, i.e. \textit{Aremonia agrimonoides (L.) DC.} and \textit{Potentilla alba L.}. In the previous clustering related to the original graph $G^P_1$, \textit{Rosaceae} were already split into two different clusters (cluster $1$ with $66$ species, and cluster $2$ with $54$ species).

The same approach was followed for the other two subgraphs corresponding to $G^P_1$ filtered version by $w_{ij} > 2$ and $w_{ij} > 3$ (not shown). The species sharing $3$ or $4$ morphological properties were retained as vertices in the network. In this case, the number of analyzed species drastically reduced to the $13 \%$ and $3.2 \%$ of the $D^{3}$ total amount of species. Thus, communities detection on such a highly reduced dataset had to be intended as a merely quantitative investigation. The most relevant insight confirmed previous result: \textit{Poaceae} family survived severe filtering, and they gathered in two different ways. Some \textit{Poaceae} species were grouped on the basis of three morphological properties, mainly: nutrient, elongated, and flat diaspore type. Some other species, usually found in the same community embedding \textit{Rosaceae} species, also showed mucilaginous diaspore surfaces. 

We can conclude that the high family heterogeneity in each cluster survives the edges-weight based filtering: diaspore morphology seems not to be a good classifier, and further analysis on different datasets are required.

\subsubsection{Fruit-based graph}
Communities detection results are summarized in Tab~\ref{tab:test2_species_fam} while a graphical view is provided in Fig.~\ref{fig5} where eight giant components are revealed. The detected clusters are clearly separated one from each other, and the vertices (plants species) are fully-connected inside each community. In other terms, the plants belonging to a cluster all share a single precise property. As for previous cases, no particular homogeneity in terms of family composition is observed (more information in the Supplementary Information). 
\begin{table}
\centering
\resizebox{0.4\textwidth}{!}{%
\begin{tabular}{|c|c|c|c|c|} 
\hline
\hline
\textbf{cluster} & \textbf{species} & \textbf{$\%$} & \textbf{families} & \textbf{$\%$}\\
\hline
\hline
1 & $1426$ & $53.57 \%$ & $47$ & $42.3 \%$\\
2 & $593$ & $22.28 \%$ & $42$ & $37.83 \%$\\
3 & $326$ & $12.25 \%$ & $24$ & $21.62 \%$\\
4 & $149$ & $5.6 \%$ & $30$ & $27.02 \%$\\
5 & $143$ & $5.37 \%$ & $11$ & $9.9 \%$\\
6 & $13$ & $0.49 \%$ & $3$ & $2.7 \%$\\
7 & $10$ & $0.38 \%$ & $5$ & $4.5 \%$\\
8 & $2$ & $0.08 \%$ & $1$ & $0.9 \%$\\
\hline
\hline
\end{tabular}}
\caption{$G^P_{2}(N,E)$ clusters composition on the basis of fruit typology categorical traits. Species percentage is referred to the relative amount of species inside each cluster with respect to the total number of species present in the database ($2,662$). Families percentage is referred to the total number of families ($111$) present in the dataset. The majority of species belong to the first three clusters, which are also the most heterogeneous in terms of families composition.}
\label{tab:test2_species_fam}
\end{table}
\subsubsection{Graph of properties, $G^F$ from diaspore morphology.}\label{sec:originalGraph}
Similarly to what has done so far we also considered the second projection giving the graph of features shown in  Fig.~\ref{fig6}. Such graph $G^F$ is composed by $N=8$ vertex and $E=15$ edges. Two nodes are completely isolated, and they correspond to properties \textbf{other specialization} and \textbf{no specialization}, in agreement with the previous findings (see Fig.~\ref{fig3} (panel A), clusters $3$ (cyan) and $5$ (red)) looking like isolated components of the graph, that is to say showing  properties that do not share any other property with the other species. The dispersal of plants characterized by such properties, also not sharing any other properties with other species, may be not crucially linked to seed or fruit morphology (and typology). Edges thickness is proportional to the number of common plants sharing the two properties connected by that link. In that sense, the elongated and flat appendages properties are common to a huge number of species. More in detail, the properties \textbf{flat-elongated},  \textbf{flat-nutrient}, \textbf{hooked-elongated}, \textbf{elongated-nutrient} share several species between them, respectively $323$, $277$, $272$, and $219$, and they have to be considered aggregative properties over the set of morphological seeds properties.

\section{Discussion} \label{sec:discussion}
Plants diaspore morphological features have been analysed in order to classify the various species. Data have been extracted from the $D^{3}$ Dispersal and Diaspore Database\cite{hintze2013d3}, developed as a partial solution to the gap about dispersal-related traits of plant species. In this paper we applied various quantitative measures, based on Complex Network Theory,  in order to measure effective similarities between various species.

In particular we applied different communities detection algorithms 
\begin{itemize}
\item
to inspect plants species with the final goal to underline salient structures characterising our data; 
\item 
to identify the degree of similarity among the different species; 
\item to organise data in smaller structures and to gain insight into general hypothesis and properties of the whole dataset.
\end{itemize}
At a first glance, diaspores morphology did not turn out to be a good classification parameter for species. Indeed, different species share more than one common property, and each community show a huge heterogeneity in terms of family composition. An explanation of this fact is that during their evolution plants were subjected to a strong selective pressure in order to colonise suitable habitats, mostly throughout the dispersal of seeds. To solve this problem, plants converged in the production of secondary structures such as plumes, samaras, hooks, wings, aerenchimas and mucilagines. Such convergent evolution determines that very similar solutions are found in species belonging to distant families. This is in accord to our results, where very different and genetically unrelated plants cluster in stable groups.
We observed the same behaviour also after a severe filtering that was applied on plants graph. Complex networks analysis main results in terms of basic quantities have been confirmed after pruning by edges weight, that is by removing species which shared a small number of properties. 

On the other hand, species can be classified by their fruit topology, which prove to be a good categorical trait. A first explanation is that probably the selection did not push enough plants to provide convergent solutions for the environment where they lived. In the same spirit we intend in the future to do further analysis on the other features provided by $D^{3}$ Dispersal and Diaspore Database, such as diaspore typology, exposure of diaspores, heterodiaspory to improve the present findings. 
In conclusion, complex networks analysis seems to be an advantageous tool to investigate plants relationships related to morphological features. We believe that a similar approach may be applied with success to the study of many other fields of plant science, such as plant ecology, phytosociology and plant communication. 
\section{Materials and Methods}\label{sec:methods}

\subsection{Data}
Data are collected in the $D^{3}$ Dispersal and Diaspore Database\cite{hintze2013d3} available at website \url{http://www.seed-dispersal.info/}.
$D^{3}$ database is developed as a partial solution to the lack of knowledge about dispersal/related traits of plant species, with the aim to simplify traditional ecological and evolutionary analysis. Currently the database provides several information related to seed dispersal of plant species, such as empirical studies, functional and heritable traits, dispersal units image analysis and ranking indices (i.e. parameters which quantify the adaptation of a species to certain seed dispersal mode, in relation to a larger species set). More than $5,000$ plant species are reported.  Available raw data are mainly provided by DIASPORUS\cite{bonn2000diasporus}, BIOPOP\cite{poschlod2003biopop} and LEDA\cite{kleyer2008leda} databases of plants traits. Here we focused on the well documented $2,662$ Central European taxa, by exploiting the detailed ecomorphological categorizations of the diaspore and fruit, as well as information on prevailing dispersal modes. For every species we took into account diaspore morphology and fruit topology. 
\subsubsection{Diaspore Morphology} 
Morphology was treated technically as a set of binary traits. During the first test, eight features were taken into account for the categorization of diaspore morphology (See Supplementary information for more details):(1) nutrients:
(2) elongated body; (3) hooked body; (4) flat/wings; (5) ballo/aerenchym; (6) mucilaginous; (7) none of the above: diaspores without any of the above mentioned  specializations; (8) vegetative specialization.
Such categorization scheme was inspired by the LEDA approach \cite{kleyer2008leda}. However, diaspore morphology represents an original dataset, which was derived either from visual inspection of the diaspores and respective images, or from an intensive and web research.
\subsubsection{Fruit Morphology} 
Fruit typology is a categorical trait which describes those ecological characteristics of
the fruit which are related to seed dispersal. In the following analysis five categorization of ecological fruit types were taken into account. Fruit typology was categorized by visual inspection of fruits or respective images in addition to an intensive literature and web research\cite{bojnansky2007atlas}. Schematically (more detail on the Supplementary Information) they are divided into (1) indehiscent fruit: the pericarp is not opening during ripening; the above is further divided in (1a) non-fleshy; (1b) fleshy fruit; (1c) pepo. (2) dehiscent fruit: the pericarp opens during ripening; further divided in (2a) fruit with upright aperture; (2b) fruit with lateral aperture. (3) explosive release; (4)  gymnosperme type; (5) not applicable: reserved for the following species (5a) sterile hybrid (e.g. \textit{Betula x aurata}); (5b) for vegetative diaspore types.

\subsection{Building the graph: projection in the space of plants/features}
From the data written in the form of a bipartite graph (where every species $N$ is connected to its features) we obtain two different projection graphs with the procedure shown in Fig.~\ref{fig1}. Once a bipartite graph is built, it can also be described by a matrix $A(p,f)$ whose element $a_{ij}$ is $1$ if plant $p$ has the feature $f$.  The most immediate way to measure correlation between species is counting how many seeds features a couple of species share and similarly how many plants share the same couple of seeds features. 
In formulas, this corresponds to  consider the matrix of species $P(p,p)=AA^T$ and the matrix of seeds features $F(f,f)=A^TA$.
In detail  we focused on the graph having as nodes the different plants, i.e. on the {\em Plants graph} $G^P(N,E)$ where  edges weights were proportional to the number of common features shared by a couple of plants (this could be diaspora-based or fruit-based). Second, in order to catch the predominant properties in terms of seeds dispersal, we analysed the second bipartite projection, i.e. the {\em Features graph}, $G^F(N,E)$, whose nodes represented the different diaspore morphological traits taken into account. In that case edges weights were proportional to the number of plants sharing the same feature. Both a network metrics analysis, and a basic cluster analysis were performed to obtain an alternative classification of plants. 

\subsection{Basic network analysis}
As regards network analysis, we computed some global and local basic metrics, described hereafter.
\begin{itemize}
\item \textit{Graph density} is defined as the ratio between the numbers of existing edges and the possible number of edges, in a $N$-size network it is given by $(\frac{2E}{N(N-1)})$. 
\item \textit{Network clustering coefficient} is the overall measure of clustering in a undirected graph in terms of probability that the adjacent vertices of a vertex are connected. More intuitively, global clustering coefficient is simply the ratio of the triangles and the connected triples in the graph. The corresponding local metric is the \textit{local clustering coefficient}, which is the tendency among two vertices to be connected if they share a mutual neighbour. In this analysis we used a local vertex-level quantity\cite{barrat2004architecture} defined in Eq.~\eqref{eq:weighted_C}: 
\begin{equation}\label{eq:weighted_C}
c_{i}^{w} = \frac{1}{s_{i}(k_{i}-1)} \sum_{jh} \frac{(w_{ij}+w_{ih})}{2}a_{ij}a_{ih}a_{jh},
\end{equation}
The normalization factor $\frac{1}{s_{i}(k_{i}-1)}$ accounts for the weight of each edge times the maximum possible
number of triplets in which it may participate, and it ensures that $0 \leq c_{i}^{w} \leq 1$.
That metric combines the topological information with the weight distribution of the network, and it is a measure of the local cohesiveness grounding on the importance of the clustered structure on the basis of the amount of interaction intensity actually found on the local triplets\cite{barrat2004architecture}.
\item \textit{Network strength} ($s$) is obtained by summing up the edge weights of the adjacent edges for each vertex\cite{barrat2004architecture}. That metric is a more significant measure of the network properties in terms of the actual weights, and is obtained by extending the definition of \textit{vertex degree} $k_{i} = \sum_{j}a_{ij}$, with $a_{ij}$ elements of the network adjacent matrix $\textbf{A}$. In formulas $s_{i} = \sum_{j = 1}^{N}a_{ij}w_{if}$.
\end{itemize}

\subsection{Grouping plants from graph: communities detection analysis} \label{sec:appendix_cd}
Communities detection aims essentially at determine a finite set of categories (clusters or communities) able to describe a data set, according to similarities among its objects\cite{campello2007fuzzy}. More in general, hierarchy is a central organising principle of complex networks, able to offer insight into many complex network phenomena\cite{clauset2008hierarchical}. 

In the present work we adopted the following method: 
\begin{itemize}
\item {\bf Fast greedy (FG)} hierarchical agglomeration algorithm\cite{clauset2004finding} is a faster version of the previous greedy optimisation of modularity\cite{newman2004finding}. FG gives identical results in terms of found communities. However, by exploiting some shortcuts in the optimisation problem and using more sophisticated data structures, it runs far more quickly, in time $O(md\log n)$, where $d$ is the depth of the ``dendrogram'' describing the network community structure.
\item {\bf Walktrap community finding algorithm (WT)} finds densely connected subgraphs from a undirected locally dense graph \textit{via} random walks. The basic idea is that short random walks tend to stay in the same community\cite{pons2005computing}. Starting from this point, $WT$ is a measure of similarities between vertices based on random walks, which captures well the community structure in a network, working at various scales. Computation is efficient and the method can be used in an agglomerative algorithm to compute efficiently the community structure of a network.
\item {\bf Louvain or Blondel method (BL)} \cite{blondel2008fast} to uncover modular communities in large networks requiring a coarse-grained description. \textit{Louvain} method ($BL$) is an heuristic approach based on the optimisation of the modularity parameter ($Q$) to infer hierarchical organization. Modularity (Eq.~\eqref{eq:Modularity}) measures the strength of a network division into modules\cite{newman2004finding,newman2004fast}, as it follows:
\begin{equation}
\label{eq:Modularity}
Q = \frac{1}{2m} \sum_{vw} \left[A_{vw} - \frac{k_{v}k_{w}}{ \left(2m \right)} \right]\delta\left(c_{v}, c_{w}\right) = \sum^{c}_{i=1}(e_{ii} - a^{2}_{i}),
\end{equation}
where, $e_{ii}$ is the fraction of edges which connect vertices both lying in the same community $i$, and $a_{i}$ is the fraction of ends of edges that connect vertices in community $i$, in formulas: $e_{ii} = \frac{1}{2m}\sum_{vw} \left[A_{vw} \delta\left(c_{v}, c_{w}\right) \right]$, and $a_{i} = \frac{k_{i}}{2m}=\sum_{i}e_{ij}$; $\textbf{A}$ is the adjacent matrix for the network; $c$ the number of communities; $k_{i} = \sum_{w} A_{vw}$ the degree of the vertex-$i$, $n$ and $m = \frac{1}{2}\sum_{vw} A_{vw}$ the number of graph vertices and edges, respectively. \textit{Delta} function, $\delta(i,j)$, is $1$ if $i=j$, and 0 otherwise.	
\item {\bf Label propagation (LP)} community detection method is a fast, nearly linear time algorithm for detecting community structure in networks\cite{raghavan2007near}. Vertices are initialised with a unique label and, at every step, each node adopts the label that most of its neighbours currently have, that is by a process similar to an `updating by majority voting' in the neighbourhood of the vertex. Moreover, $LP$ uses the network structure alone to run, without requiring neither optimisation of a predefined objective function nor \textit{a-priori} information about the communities, thus overcoming the usual big limitation of having communities which are implicitly defined by the specific algorithm adopted, without an explicit definition. In this iterative process densely connected groups of nodes form a consensus on a unique label to form communities.
\end{itemize}

\section*{Acknowledgments}
The authors acknowledge support from EU FET Open Project PLEASED nr. 296582. 
GV and GC also acknowledge EU FET Integrated Project MULTIPLEX nr. 317532. 
SM and GC are particularly indebted with C. Tomei for many interesting discussions in his Lab.

\section*{Author contributions statement}
All authors contributed equally to the analysis of the dataset and to the
interpretation of the results of this analysis, both from the point of view of
Network Theory as well as in terms of biological implications.
They also contributed equally to the writing and reviewing of the manuscript.

\section*{Additional information}
\textbf{Competing financial interests.} The authors declare no competing financial interests.

\newpage

\newpage
\section{Supplementary Information: Networks of plants: how to measure similarity in vegetable species}

\flushbottom
\maketitle
\thispagestyle{empty}

\subsection{Families present in the Graph $G^P_1$}
Here we show the structure of families present in the first projection graph where common features are diaspora-based.
\begin{table}
\centering
\resizebox{0.6\textwidth}{!}{%
\begin{tabular}{|c|c|c|c|c|c|} 
\hline
\hline
\textbf{Fam.} & \textbf{cl 1} & \textbf{Fam.} &  \textbf{cl 2} &\textbf{Fam.} & \textbf{cl 3} \\
\hline
\hline
\color{red}Poaceae & \color{red}231 & \color{blue}Asteraceae & \color{blue}279  & Caryophyllaceae &  86\\
Fabaceae & 116  & \color{darkgreen}Cyperaceae	& \color{darkgreen}134  & Orchidaceae     & 61\\
\color{cerise}Rosaceae & \color{cerise}66   & Ranunculaceae	& 66 & Orobanchaceae   &  48\\
Plantaginaceae	& 30  & {\color{cerise}Rosaceae} & {\color{cerise}54} & Brassicaceae &  43\\
Polygonaceae & 28 & Apiaceae & 36 & \color{blue}Asteraceae & \color{blue}41\\
Violaceae &	23 & Amaranthaceae	& 34 &  Apiaceae  & 38\\
Apiaceae &	22 & Salicaceae & 32 & Rubiaceae &  30 \\
Amaranthaceae	& 20     & Caprifoliaceae	& 27& Primulaceae &  29 \\
Juncaceae &	17 & Potamogetonaceae	& 23 & Campanulaceae &  28\\
Papaveraceae &	17 & Lamiaceae	& 21 & Saxifragaceae &  27 \\
Boraginaceae &	16 & Onagraceae	& 21 &  Lamiaceae &  26 \\
Lamiaceae &	16  & Brassicaceae	& 20 & Crassulaceae &  22 \\
Orobanchaceae &	16 & Boraginaceae	& 18     & Gentianaceae &  22 \\
Caryophyllaceae	 & 14  & Rubiaceae	& 9 &   Rosaceae &  22 \\
Ericaceae &	14 & Typhaceae & 8 & Plantaginaceae &  19 \\
Betulaceae & 9  & Geraniaceae & 7 & Amaryllidaceae & 18 \\
Caprifoliaceae & 9 & Plumbaginaceae	& 7  &  Ericaceae  & 18 \\
Pinaceae &	9 & Alismataceae	& 6 & Scrophulariaceae &  14 \\
Santalaceae	& 9 & Caryophyllaceae	& 6 & Convolvulaceae &  12 \\
Solanaceae &	9   & Fabaceae	& 6 &  Ranunculaceae  & 12\\
Asparagaceae &	8    & Urticaceae	& 6 &   Asparagaceae &  11\\
\hline
\hline
\end{tabular}}
\caption{Major families found in $G_{1,P}(N,E)$ clusters $1$, $2$, and $3$ (the largest ones) by modularity (BL) algorithm, and the corresponding number of species belonging to them.}
\label{tab:largest_clusters}
\end{table}

Such  communities are not homogeneous in terms of family composition (see Fig.~\ref{fig3}). Hereafter each cluster composition is summarized, together with the morphological properties that the element families share each other. Notice that one property can be shared by more than a single species in the same cluster, since diaspore morphological features are not mutually exclusive. 

\begin{itemize}
\item {\colorbox{yellow}{cluster 1}}: $884$ species ($33.21 \%$ of database $D^{3}$ total species); prevailing families: \textit{Poaceae}, \textit{Fabaceae}, \textit{Rosaceae}, \textit{Plantaginaceae}, \textit{Polygonaceae} (Tab.~\ref{tab:largest_clusters}, first column). $709$ species have \textbf{nutrient} diaspores, followed by $447$ showing \textbf{flat/wings} diaspore morphology; $204$ times is encountered the \textbf{elongated} feature.
\item {\colorbox{fuchsia}{cluster 2}}: $858$ species ($32.23 \%$) dominant families: \textit{Asteraceae}, \textit{Cyperaceae}, \textit{Ranunculaceae}, \textit{Rosaceae}, \textit{Apiaceae}, \textit{Apiaceae}, \textit{Amaranthaceae}, \textit{Salicaceae}, \textit{Caprifoliaceae} (Tab.~\ref{tab:largest_clusters}, second column). The vast majority of the species ($782$) show \textbf{elongated} diaspore trait; other common observed properties are: \textbf{hooked} ($220$), \textbf{ballo/aerenchym} ($224$), and \textbf{flat/wings} ($140$).
\item {\colorbox{aqua}{cluster 3}}: $753$ species ($28.29 \%$), sharing property \textit{no specialization}. Notwithstanding its big dimensions, that cluster is a completely isolated component robust to changes in clustering algorithms. The leading families belonging to cluster cyan are summarized in Tab.~\ref{tab:largest_clusters} (third column). They all share the same \textbf{no specialization} property concerning diaspore morphology. That category refers to species whose diaspores can have either a structured surface and no further appendages or specializations (e.g. many \textit{Caryophyllaceae}), or a smooth surface and no further appendages or specializations (e.g. many \textit{Brassicaceae}). Table~\ref{tab:largest_clusters} confirms that behaviour, since \textit{Caryophyllaceae} and  \textit{Brassicaceae} are two of the most numerous families with $86$ and $43$ species each respectively, besides \textit{Orchidaceae} ($61$) and \textit{Orobanchaceae} ($48$). 
\item {\colorbox{green}{cluster 4}}: $157$ species ($5.9 \%$); prevailing families: \textit{Brassicaceae}, \textit{Juncaceae}, \textit{Plantaginaceae}, \textit{Asteraceae}, \textit{Lamiaceae}. All these species share \textbf{mucilaginous} diaspore property. 
\item {\colorbox{red}{\textbf{\color{white}{cluster 5}}}}: 9 plants species belonging to \textit{Hydrocharitaceae}, \textit{Brassicaceae}, \textit{Polygonaceae}, and \textit{Araceae} families. They all show \textbf{other specialization} concerning diaspore morphology. More in detail, 7 out of 9 are aquatic plants (5 species of \textit{Hydrocharitaceae} and 2 of \textit{Araceae} family); 1 species belongs to \textit{Brassicaceae} and 1 to \textit{Polygonaceae}. The 5 species of \textit{Hydrocharitaceae} are strictly related: like other \textit{Hydrocharitaceae}, they are aquatic plants that release their diaspore in water and that, conversely to other plants of the same family, have seeds with very low nutrients content; more, they do not set seeds regularly, preferring asexual reproduction; in both cases (sexual or asexual reproduction) water movements allow the dispersal; the 2 other aquatic (\textit{Araceae}) also prefer asexual reproduction; having no or little roots, the whole plants can float and disperse; the species belonging to the family of \textit{Brassicaceae} has dehishent fruits; finally, the species of \textit{Polygonaceae} rarely produces viable seeds and reproduction is normally asexual (by bulbils)
\item {\colorbox{blue}{\textbf{\color{white}{cluster 6}}}}: 1 isolated plant, \textit{X Calammophila baltica Brand (Poaceae)} which doesn't show any of the used morphological properties with the other species. 
\end{itemize}

Table~\ref{tab:largest_clusters_g1_wlarg1} refers to the communities detection results after pruning the graph. 
\begin{table}
\centering
\resizebox{0.8\textwidth}{!}{%
\begin{tabular}{|c|c|c|c|c|c|c|c|c} 
\hline
\hline
\textbf{Fam.} & \textbf{cl 1} & \textbf{Fam.} & \textbf{cl 2} &  \textbf{Fam.} &  \textbf{cl 3} & \textbf{Fam.} & \textbf{cl 4}\\
\hline
\hline
{\color{red}Poaceae} & {\color{red}228} & {\color{darkgreen}Cyperaceae} & {\color{darkgreen}89} & {\color{cerise}Rosaceae} & {\color{cerise}23} & Potamogetonaceae & 20\\
Juncaceae & 14  & {\color{cerise}Rosaceae} & {\color{cerise}48} & \color{darkgreen}Cyperaceae & \color{darkgreen}6 & Plantaginaceae	& 19\\
Fabaceae &	11  & Ranunculaceae   & 42   & Fabaceae & 3 & Amaranthaceae	& 12\\
Santalaceae	& 9 & \color{blue}Asteraceae & \color{blue}29 & Nymphaeaceae & 2 & \color{blue}Asteraceae & \color{blue}7\\
Caprifoliaceae & 8 & Apiaceae & 26 & Amaranthaceae & 1 & Brassicaceae	& 7\\
Pinaceae &	8    &  Lamiaceae  & 17 & Araceae & 1 & Juncaceae	& 3\\
Polygalaceae &	8    & Boraginaceae &  16 & Juncaginaceae & 1 & Lamiaceae & 1\\
Amaranthaceae	& 7     & Caprifoliaceae &  16 &   &     & 	& \\
Plumbaginaceae &	7    & Polygonaceae &  10 &   &     & 	& \\
Lamiaceae &	6    & Rubiaceae &  9 &   &     & 	& \\
Orobanchaceae &	6      &  Geraniaceae &  6 &   &     & 	& \\
Sapindaceae &	6     & Alismataceae &  5&   &     & 	& \\
Plantaginaceae &	4     & Typhaceae &  4 &   &     & 	& \\
\hline
\hline 
\end{tabular}}
\caption{Families belonging to each of the four clusters identified by communities detection. Graph $G^P_1(N,E)$ is filtered by edges weight $w_{ij} > 1$.}
\label{tab:largest_clusters_g1_wlarg1}
\end{table}
Again, detected communities are not homogeneous in terms of family composition. Anyway, more correspondences can be observed between the two panels of Fig~\ref{fig4}. Red and cyan clusters, for example, are less heterogeneous, being composed by \textit{Poaceae} 
and \textit{Rosaceae} families, respectively (white and cerise dots in the right panel). 

It follows a brief description of the four clusters identified by BL method.
\begin{itemize}
\item {\colorbox{red}{\textbf{\color{white}{cluster 1}}}}: $352$ species ($43.84 \%$ of database $D^{3}$ total species); \textit{Poaceae} with $228$ species are clearly the prevailing family: see white nodes in the right panel of Fig.~\ref{fig4}. They are followed by \textit{Juncaceae} ($14$ plants), \textit{Fabaceae}, \textit{Santalaceae}, \textit{Caprifoliaceae}, \textit{Pinaceae}. 
\\ All these species share that common properties: \textbf{nutrients} ($315$), \textbf{flat/wings} ($312$), \textbf{elongated} ($240$). They do not show (almost most of them)  \textbf{ballo/aerenchyms} and \textbf{mucilaginous} surfaces;
\item {\colorbox{fuchsia}{cluster 2}}: $345$ species ($42.96 \%$); dominant families: \textit{Cyperaceae} ($89$), \textit{Rosaceae} ($48$), \textit{Ranunculaceae}($42$), \textit{Asteraceae}($29$). \textit{Cyperaceae} are visible as red dots in Fig.~\ref{fig4} (panel B) in the position corresponding to violet cluster of left panel.
That cluster embeds species joined by \textbf{elongated} ($317$) and \textbf{hooked} ($211$) diaspores shape. \textbf{Ballo/aerenchyms} and \textbf{flat/wings} are shared by $175$ and $112$ species, respectively. Just $4$ species shows \textbf{mucilaginous} surfaces;
\item {\colorbox{aqua}{cluster 3}}: $37$ species ($8.95 \%$); \textit{Rosaceae} family dominates with $23$ species, visible as cerise vertices in Fig.~\ref{fig4} (panel B) in the position corresponding to cyan cluster in the left panel. Almost all of them share clearly two properties: \textbf{nutrients} and \textbf{ballo/aerenchyms} surfaces;
\item {\colorbox{green}{cluster 4}}: $69$ species ($4.61 \%$), dominated by those belonging to \textit{Potamogetonaceae} ($20$), \textit{Plantaginaceae} ($19$), and \textit{Amaranthaceae} ($12$) families. All the species have \textbf{mucilaginous} surfaces, some of them show \textbf{flat} diaspores ($39$), in particular species belonging to \textit{Plantaginaceae} and \textit{Juncaceae} families; other individuals show \textbf{elongated} diaspore ($41$), especially \textit{Amaranthaceae}, \textit{Asteraceae}, \textit{Potamogetonaceae}.
\end{itemize}

\subsection{Graph of plants $G^P_2(N,E)$ from fruit typology.}\label{sec:fruitTypoGraph}

\begin{figure}
\includegraphics[width=1\textwidth]{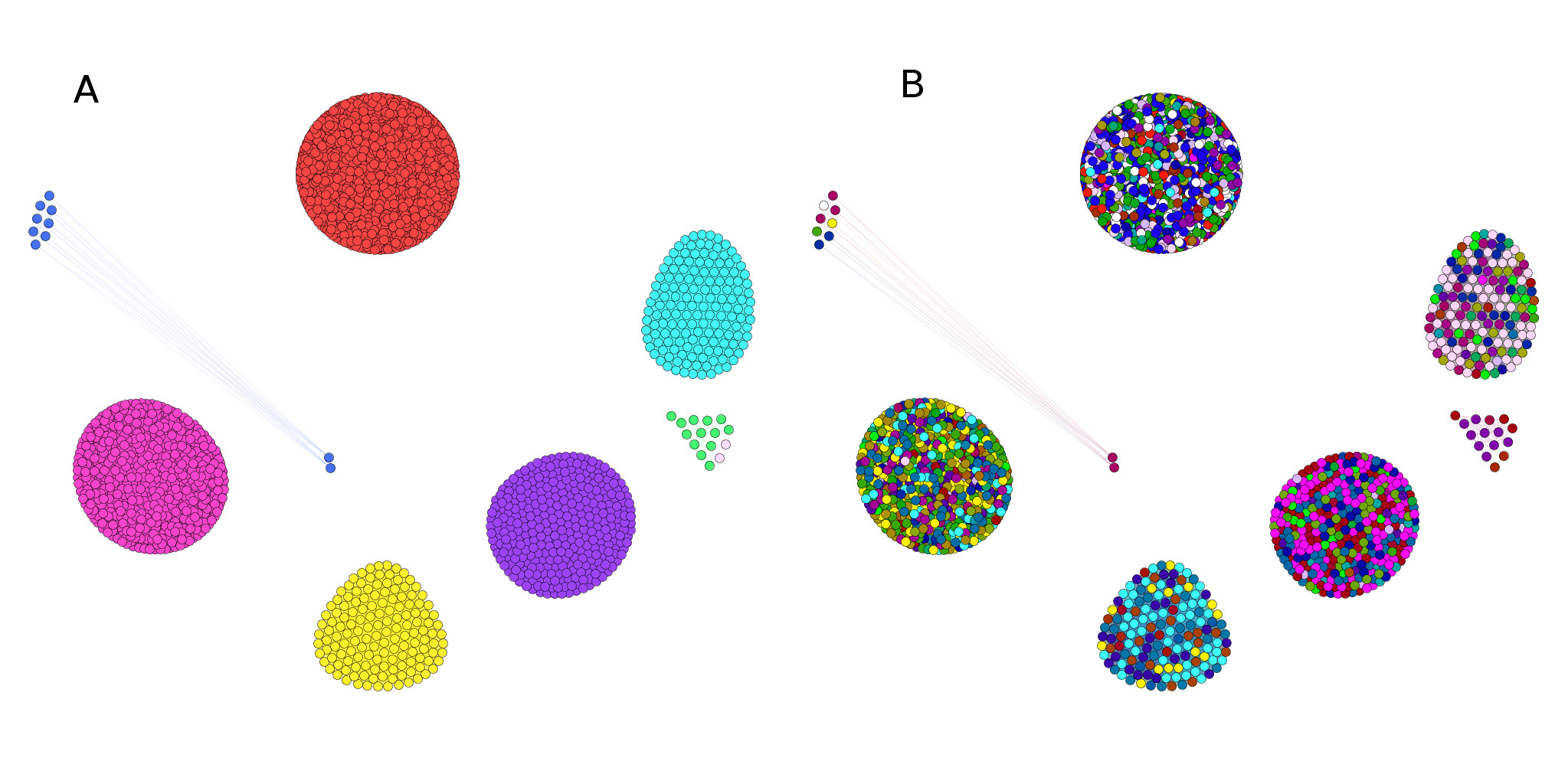} 
\caption{\textbf{Fruit typology graph communities.} $G^2_P(N,E)$ communities detection by modularity method (BL). Only edges with weight $w_{ij} = 1$ are present. Eight isolated communities are detected ( panel A), and the corresponding families composition is displayed (panel B). Clearly each cluster is highly heterogeneous in terms of families composition, but not in terms of shared properties between the species belonging to each cluster. A single fruit topological property, in fact, is associated to each cluster and species. Main families are visible: \textit{Poaceae} (white), \textit{Asteraceae} (blue), \textit{Cyperaceae} (red), \textit{Rosaceae} (cerise), \textit{Fabaceae} (cyan), \textit{Caryophyllaceae} (fuchsia).} 
\label{fig5}
\end{figure}

\begin{figure}
\centering
\includegraphics[width=0.5\textwidth]{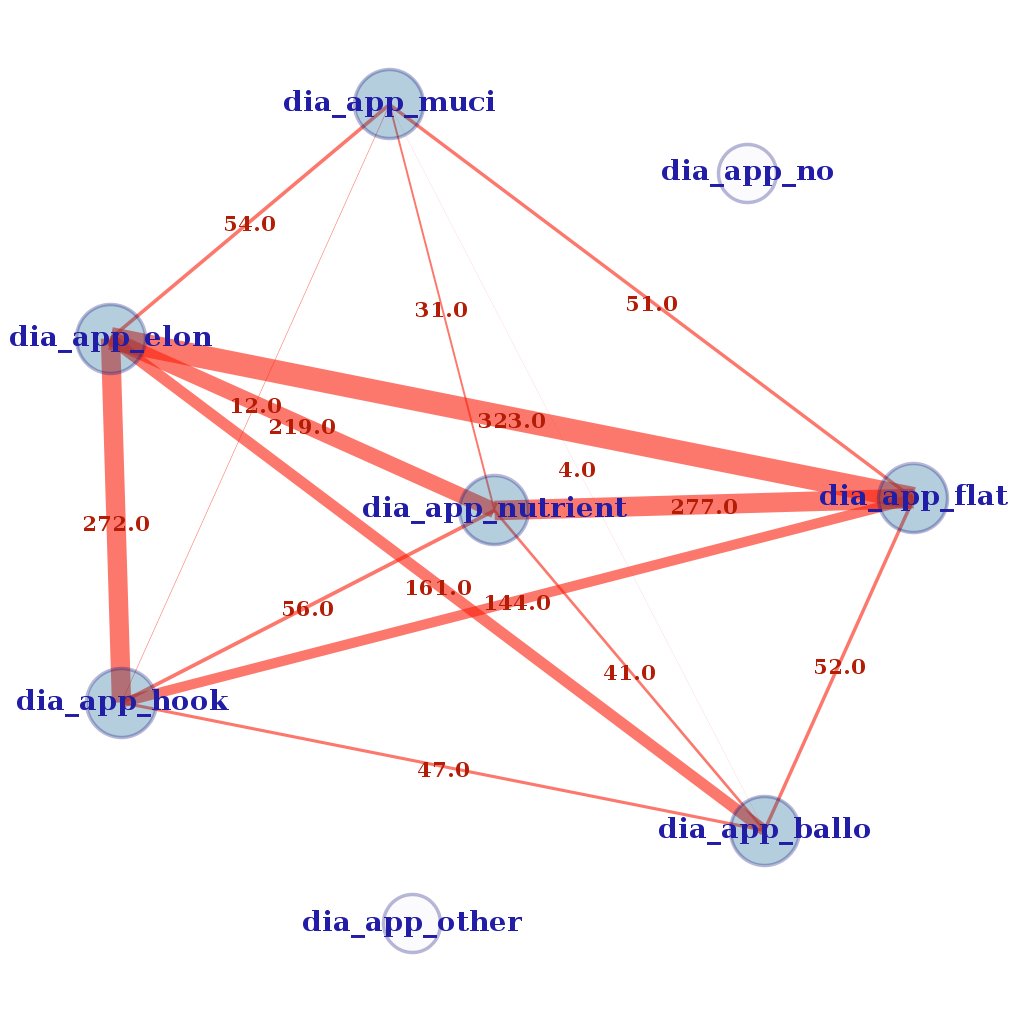} 
\caption{\textbf{Morphological properties network.} The following figure shows the second bipartite projection obtained by connecting features between them ($G^F(N,E)$) on the basis of how many plants share a given property. $N=8$ vertices (blue circles) and $E=15$ edges (red lines). 
Two nodes are isolated: they correspond to properties \textit{other specializations} and \textit{no specialization}. Thicker edges correspond to edges with higher weights (whose values are reported near each node), while nodes dimensions are proportional to the node degree (larger for higher degrees). It follows that properties \textit{flat-elongated},  \textit{flat-nutrient}, \textit{hooked-elongated}, \textit{elongated-nutrient} share more species between them, respectively $323$, $277$, $272$, and $219.$}
\label{fig6}
\end{figure}

As regards the fruit-based graph we have here a short description of the detected communities, together with the main families belonging to them (Tab.~\ref{tab:largest_clusters_c2}), and the topological properties of the corresponding species fruits. The graph is shown in Fig.~\ref{fig5}.
\begin{itemize}
\item {\colorbox{red}{cluster 1}}: $1426$ species belonging to $47$ different families, mainly to \textit{Asteraceae} ($341$), \textit{Poaceae} ($231$), and \textit{Cyperaceae} ($150$), \textit{Apiaceae} ($95$), and \textit{Rosaceae} ($84$). All these species are characterized by \textbf{non fleshy indehiscent fruit} (hard or woody pericarp).

\item {\colorbox{fuchsia}{cluster 2}}: $593$ species, mainly \textit{Brassicaceae}($116$), \textit{Orchidaceae}($61$), \textit{Orobanchaceae}($58$), \textit{Plantaginaceae}($49$), \textit{Fabaceae}($48$), all showing \textbf{dehiscent fruit with lateral aperture}, i.e. a configuration allowing seeds to be released faster. 

\item {\colorbox{darkorchid}{cluster 3}}: $326$ species, especially \textit{Caryophyllaceae} ($99$), \textit{Juncaceae}($43$), \textit{Primulaceae}($37$), \textit{Saxifragaceae}($27$), \textit{Crassulaceae}($22$). That species are characterized by \textbf{dehiscent fruit with upright aperture}, allowing seeds to stay a longer time in the open fruit.

\item {\colorbox{aqua}{cluster 4}}: $149$ species being part of \textit{Rosaceae} ($56$), \textit{Ericaceae} ($11$), \textit{Solanaceae} ($9$), and \textit{Asparagaceae} ($7$) families, showing \textbf{fleshy indescent fruit}.

\item {\colorbox{yellow}{cluster 5}}: $143$ species mainly belonging to \textit{Fabaceae}, \textit{Euphorbiaceae}, \textit{Violaceae}, \textit{Geraniaceae}, and \textit{Brassicaceae} families, all characterized by an \textbf{explosive release mechanism}.
\item {\colorbox{green}{cluster 6}}: $13$ species subdivided as it follows: $9$ belonging to \textit{Pinaceae}, $3$ to \textit{Cupressaceae}, and $1$ to \textit{Taxaceae} families, respectively. They all share \textbf{gymnosperme} seeds with or without hull structures.
\item {\colorbox{lightblue}{cluster 7}}: $10$ species belonging to \textit{Hydrocharitaceae} ($5$), \textit{Araceae} ($2$) and some species belonging to \textit{Brassicaceae}, \textit{Poaceae}, \textit{Polygonaceae} families, mainly. All that species show \textbf{not applicable} typology of fruit, typical of those species which either do not produce diaspore or do show vegetative diaspore types.

\item {\colorbox{lavanda}{cluster 8}}: $2$ species belonging to \textit{Cucurbitaceae} family: \textit{Bryonia alba L.} and \textit{Bryonia dioica Jacq.}, both showing just \textbf{pepo} indehiscent fruit typology.
\end{itemize}

\begin{table}
\centering
\resizebox{0.7\textwidth}{!}{%
\begin{tabular}{|c|c|c|c|c|c|c|c|} 
\hline
\hline
\textbf{Fam.} & \textbf{cl 1} & \textbf{Fam.} &  \textbf{cl 2} &\textbf{Fam.} & \textbf{cl 3} & \textbf{Fam.} & \textbf{cl 4} \\
\hline 
\hline
\color{blue}Asteraceae & \color{blue}341 & Brassicaceae & 116  & Caryophyllaceae &  99 & \color{cerise}Rosaceae & \color{cerise}56\\
\color{red}Poaceae & \color{red}231  & Orchidaceae	& 61  & Juncaceae & 43 & Ericaceae & 11\\
\color{darkgreen}Cyperaceae & \color{darkgreen}150   & Orobanchaceae & 58 & Primulaceae  &  37 & Solanaceae & 9\\
Apiaceae	& 95  & Plantaginaceae & 49 & Saxifragaceae &  27 & Asparagaceae & 7\\
\color{cerise}Rosaceae & \color{cerise}84 & \color{cyan}Fabaceae & \color{cyan}48 & Crassulaceae & 22 & Caprifoliaceae & 7\\
Lamiaceae &	73 & Salicaceae	& 32 &  Amaryllidaceae  & 15 & Grossulariaceae & 7\\
Ranunculaceae &	68 & Campanulaceae & 30 & Amaranthaceae &  13 & Adoxaceae & 6\\
Amaranthaceae	& 44 & Gentianaceae	& 27 & Ericaceae &  13 & Araceae & 5\\
Boraginaceae &	43 & Onagraceae & 22 & Plantaginaceae &  11 & Rhamnaceae & 5\\
Rubiaceae &	39 & Scrophulariaceae	& 15 & Iridaceae &  7 & Thymelaeaceae & 5\\
Polygonaceae &	36 & Lentibulariaceae	& 11 &  Papaveraceae &  6 & Hydrocharitaceae & 4\\
Caprifoliaceae & 33 & Liliaceae	& 11 & Ranunculaceae &  6 & Nymphaeaceae & 4\\
Potamogetonaceae &	24 & Asparagaceae & 10     & Orobanchaceae &  5 & Cornaceae & 3\\
\color{cyan}Fabaceae & \color{cyan}23 & Hypericaceae & 10 &   Campanulaceae &  4 & Santalaceae & 3 \\
Plantaginaceae & 14 & Cistaceae & 9 & Celastraceae &  4 & Vitaceae & 2\\
Brassicaceae & 13  & Ranunculaceae & 9 & Asparagaceae & 2 & Acoraceae & 1 \\
Betulaceae & 10 & Ericaceae	& 8  &  Gentianaceae  & 2 & Amaranthaceae & 1\\
Malvaceae & 10  & Linaceae	& 8 & Linderniaceae &  2 & Amaryllidaceae & 1\\
Convolvulaceae	& 8 & Papaveraceae	& 8 & Solanaceae &  2 & Aquifoliaceae & 1\\
Typhaceae &	8   & Polygalaceae	& 8 &  Tofieldiaceae  & 2 & Araliaceae & 1\\
Fagaceae &	7    & Droseraceae	& 5 &   Butomaceae &  1 & Berberidaceae & 1\\
Plumbaginaceae &	7   & Amaryllidaceae	& 4 &  Colchicaceae &  1 & Caryophyllaceae & 1\\
Alismataceae & 6  & Convolvulaceae	& 4 & Linaceae &  1 & Dioscoreaceae & 1\\
\hline
\hline
\textbf{Fam.} & \textbf{cl 5} & \textbf{Fam.} &  \textbf{cl 6} &\textbf{Fam.} & \textbf{cl 7} & \textbf{Fam.} & \textbf{cl 8} \\
\hline 
\hline
\color{cyan}Fabaceae & \color{cyan}51 & Pinaceae & 9 & Hydrocharitaceae & 5 & Cucurbitaceae & 2\\
Euphorbiaceae & 24 & Cupressaceae & 3 & Araceae & 2 & &\\
Violaceae & 23 & Taxaceae & 1 & Brassicaceae & 1 & &\\
Geraniaceae & 18 & & & Poaceae & 1 & &\\
Brassicaceae & 14 & & & Polygonaceae & 1 &&\\
Oxalidaceae & 4 & && && &\\
Balsaminaceae & 3  & && && &\\
Montiaceae & 3  & && && &\\
Apiaceae & 1  & && && &\\
Cucurbitaceae & 1  & && && &\\
Rutaceae & 1  & && && &\\
\hline
\end{tabular}}
\caption{Families belonging to the eight clusters identified by communities detection of graph $G^P_2(N,E)$. That results are robust with respect to changes in detection algorithms.}
\label{tab:largest_clusters_c2}
\end{table}

\newpage

\end{document}